\documentclass{article}
\usepackage{spconf,amsmath,graphicx}

\usepackage{multirow}
\usepackage{float}
\usepackage{amsfonts,amssymb}
\usepackage[colorlinks,linkcolor=blue]{hyperref}    
\title{DUALITY TEMPORAL-CHANNEL-FREQUENCY ATTENTION ENHANCED SPEAKER REPRESENTATION LEARNING}
\name{Li Zhang, Qing Wang, Lei Xie$^*$ \thanks{Corresponding Author.}}
\address{Audio, Speech and Language Processing Group (ASLP@NPU), School of Computer Science, \\
 Northwestern Polytechnical University, Xi'an, China }
 
\begin{document}
%
\maketitle
\begin{abstract}
The use of channel-wise attention in CNN based speaker representation networks has achieved remarkable performance in speaker verification (SV). But these approaches do simple averaging on time and frequency feature maps before channel-wise attention learning and ignore the essential mutual interaction among temporal, channel as well as frequency scales. To address this problem, we propose the Duality Temporal-Channel-Frequency (DTCF) attention to re-calibrate the channel-wise features with aggregation of global context on temporal and frequency dimensions. Specifically, the duality attention -  time-channel (T-C) attention as well as frequency-channel (F-C) attention - aims to focus on salient regions along the T-C and F-C feature maps that may have more considerable impact on the global context, leading to more discriminative speaker representations.  We evaluate the effectiveness of the proposed DTCF attention on the CN-Celeb and VoxCeleb datasets. On the CN-Celeb evaluation set, the EER/minDCF of ResNet34-DTCF are reduced by 0.63\%/0.0718 compared with those of ResNet34-SE. On VoxCeleb1-O, VoxCeleb1-E and VoxCeleb1-H evaluation sets, the EER/minDCF of ResNet34-DTCF achieve 0.36\%/0.0263, 0.39\%/0.0382 and 0.74\%/0.0753 reductions compared with those of ResNet34-SE. 

\end{abstract}
\begin{keywords}
speaker representation, CNN, attention   
\end{keywords}
\section{Introduction}
\label{sec:intro}
Speaker verification~(SV) is to decide to accept or reject test utterances according to the enrollment utterances of the claimed speaker \cite{bimbot2004tutorial}. Recently, speaker verification based on neural networks has occupied the predominance with the distinguished speaker representation learning ability \cite{reynolds2002overview,bai2021speaker}. In general, an SV system based on neural networks consists of a speaker representation extractor and a supervised classification loss \cite{wan2018generalized,liu2021effective,zhou2020dynamic}. Specifically, the speaker representation extractor includes a frame-level feature extractor and an utterance-level speaker embedding aggregator. The most vital work in SV is to construct a speaker representation extractor that can generate discriminative speaker embedding~\cite{zhou2021resnext,snyder2018x,desplanques2020ecapa}.

Currently, convolutional neural network~(CNN) is the most popular neural network topology of speaker representation extractor in speaker verification \cite{ketkar2017convolutional}. Variants of CNN-based speaker representation models have been explored because of their strong representation extracting ability. X-vector \cite{snyder2018x,kanagasundaram2019study,rouvier2021review} based on 1D convolution is a pioneer work in speaker verification. Additionally, ResNet \cite{he2016deep,rahman2018attention} based on 2D convolution has become predominant in various speaker verification challenges~\cite{zeinali2019but,heo2020clova,yao2020multi,thienpondt2021integrating}. However, convolution neural networks are not flawless. The mechanism of convolution operation is to adopt a fixed-size convolution kernel to capture the local time and frequency speaker patterns in speech. But the fixed-size kernel limits the receptive field on speech features. Thus, representation extracting ability \cite{hu2018squeeze,park2018bam,wang2018non} is constrained without the essential global context information which is also important. Meanwhile, convolution neglects the interaction among channels~(convolutional filters) as well. 

To mitigate the above problems, attention mechanism \cite{vaswani2017attention} has been introduced in ResNet and x-vector. Zhou et al. \cite{zhou2019deep} adopted channel-wise response feature attention in ResNet to enhance the speaker representation ability, which was the first work introducing Squeeze-and-Excitation~(SE)~\cite{hu2018squeeze} attention in the speaker verification. The SE attention mainly learns the inter-dependency relationships of channels. 
Its superior performance has triggered awareness of adopting a variety of attention modules in the residual block of ResNet further to promote CNN's discriminative speaker representation learning ability. Inspired by the convolutional block attention module~(CBAM) \cite{woo2018cbam} in computer vision, Sarthak and Rai \cite{yadav2020frequency} proposed the multi-path~(Channel, Temporal and Frequency respectively) attention of intermediate tensors in the residual blocks of ResNet. In addition to the SE attention, this work further introduced a time-frequency domain attention mask learned by large-size kernels. 
Xia and Hansen \cite{xia2020speaker} further proposed to combine the SE attention and non-local time-frequency attention to re-calibrate the channel and temporal-frequency feature maps. Meanwhile, some recent works \cite{desplanques2020ecapa,thienpondt2021integrating} also introduced the SE attention into TDNN~(1D convolutional neural network)~\cite{rouvier2021review} with excellent performance in speaker verification. 

Although the above SE-attention approaches have achieved superior SV performance, there is still substantial space to further improve the discriminative power of attention-guided speaker embeddings. Note that these approaches do \textit{averaging} on time and frequency feature maps before channel-wise attention learning and ignore the essential mutual interaction among temporal, channel as well as frequency scales.

In this study, we propose the Duality Temporal-Channel-Frequency~(DTCF) attention to assemble the temporal and frequency information into the channel-wise attention.
Recent studies \cite{thienpondt2021integrating,yadav2020frequency} have shown that adopting attention mechanism in frequency domain has great potential to improve the representation ability of speaker embedding models. There are important interactions between channel, temporal and frequency dimensions, which can lead to clear performance gain. 
In this paper, instead of simple averaging on time and frequency feature maps in previous studies, we specifically encode temporal and frequency information into channel-wise attention, resulting in the so-called \textit{duality} attention~--~time-channel~(T-C) attention as well as frequency-channel~(F-C) attention. The proposed duality attention mechanism aims to focus on salient regions along the T-C and F-C feature maps that may have more considerable impact on the global context, leading to more discriminative speaker representations. 
Experimental results show that our proposed attention module significantly outperforms the SE attention and other competitive attention modules. 

The most similar structure to ours is \cite{hou2021coordinate}, in which coordinate attention is introduced to mobile networks  \cite{sandler2018mobilenetv2,zhou2020rethinking} to solve computer vision problems. Specifically, the coordinate attention embeds positional information of images into the channel attention, achieving promising performance in the image segmentation task. In this paper, we introduce a similar structure, the DTCF attention to the popular ResNet in speaker verification, to explore the salient regions in time-channel and frequency-channel dimensions in speech signals, aiming to learn more discriminative speaker representation.

The rest of the paper is organized as follows. Section 2 overviews of the speaker representation extractor framework. Section 3 details our proposed attention module. Section 4 introduces the experimental setup and Section 5 summarizes the experimental results. Finally, conclusions are drawn in Section 6. 

\section{Speaker Representation Framework}
\label{sec:format}
Deep residual networks \cite{he2016deep} are widely used in image recognition and have recently been applied to speaker representation learning \cite{reynolds2002overview,chung2019delving}. In our study, the backbone of speaker representation learning network is the classical ResNet34~\cite{heo2020clova,zhang2021multi}, which has achieved good results \cite{chung2020in,torgashov2020id} on different datasets of speaker verification.

\begin{figure}[t]
    \centering
    \includegraphics{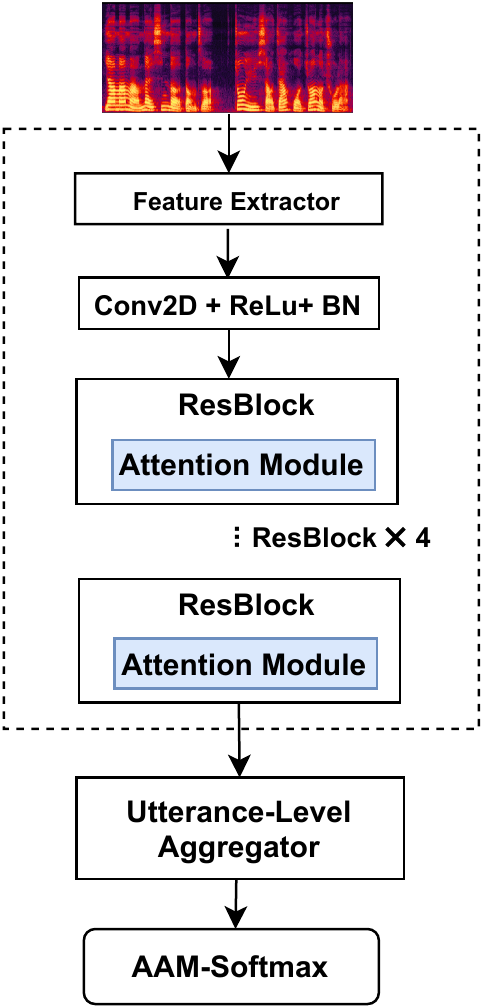}
    \caption{Overview of Speaker Representation Learning }
    \label{fig:my_label}
\end{figure}

The overview of the speaker representation learning framework is illustrated in Fig.~\ref{fig:my_label}. It mainly consists of two modules, which are speaker representation extractor and classifier respectively. The speaker representation extractor includes two sub-modules, which are a frame-level representation extractor in the dashed frame in Fig.~\ref{fig:my_label} and an utterance-level aggregator. In general, there are four residual blocks in ResNet34. The output of each residual block flows through the attention block (e.g., SE-attention) to learn the inter-dependence of channels before the skip connection~\cite{zhou2019deep}. The utterance-level aggregator encodes various length frame-level speaker representations into fixed-length utterance-level speaker representations.
We use the attention statistic pooling~(ASP)~\cite{okabe2018attentive} aggregator in the speaker representation framework. In the classification module, we adopt the Additive Angular Margin Softmax~(AAM-Sotfmax) \cite{deng2019arcface} as the loss function. 

In this paper, we mainly focus on expanding the respective field of CNN with innovative channel-wise attention as well as exploiting the global context information of time and frequency dimensions. Thus, we introduce the Duality Temporal-Channel-Frequency~(DTCF) attention mechanism after each residual block in ResNet34 to encode the temporal and frequency information into the channel-wise attention masks. Compared with other channel-wise attention modules or their variants, the DTCF attention can preserve the time and frequency information as well as learn the interactions among temporal, channel, and frequency. 
 
\section{Duality Temporal-Channel-Frequency Attention}
\label{sec:pagestyle}
\subsection{SE Attention}
\begin{figure}[htb]
 
\begin{minipage}[b]{1.0\linewidth}
  \centering
  \centerline{\includegraphics[width=8.5cm]{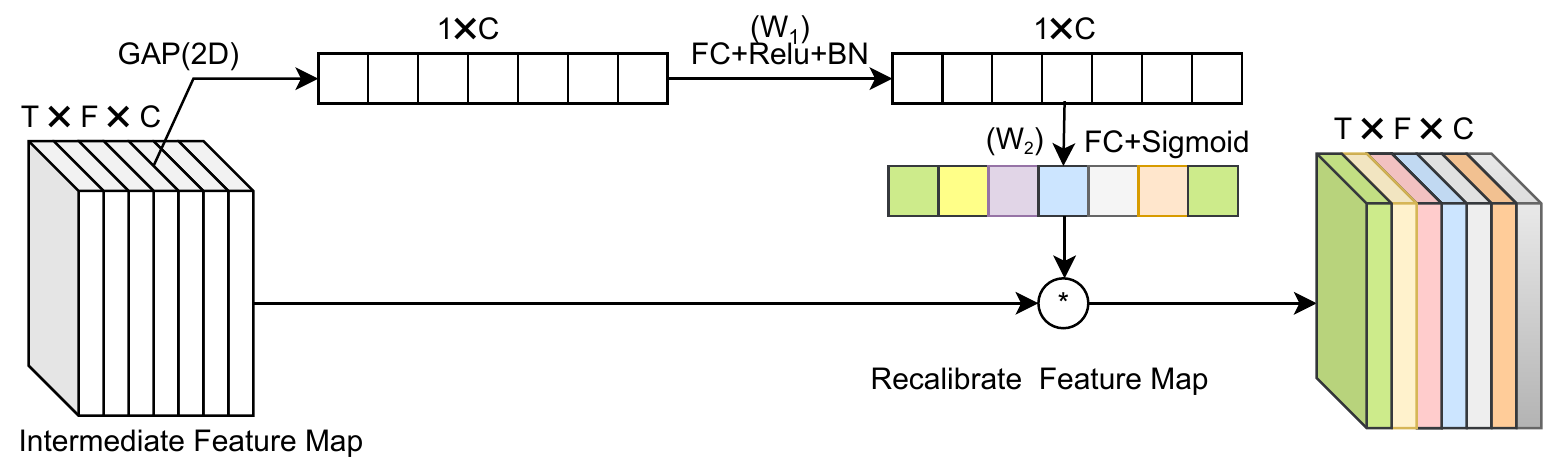}}
\end{minipage}
\caption{A Squeeze-and-Excitation~(SE) Attention Block}
\label{fig:SE}
 
\end{figure}

Channel-wise attention modules are widely embedded in ResNet to learn the inter-dependence of channels in intermediate feature maps. As shown in Fig.~\ref{fig:SE}, SE attention adopts global average pooling~(GAP) to squeeze the 3D temporal-channel-frequency tensor into a 1D channel-wise feature, then with two fully connected~(FC) layers and a sigmoid method to weigh different channel importance for boosting speaker representation learning ability. Given the intermediate tensor $X \in \mathbb{R}^ {C \times T \times F}$ is the 3D speech feature map in networks, $C$ is the number of channels (convolutional kernels), $T$ is the number of frames in the temporal domain, $F$ is the number of frequency bins. The average pooling on time and frequency dimension of SE attention \cite{hu2018squeeze} is formulated as:
\begin{equation}
\begin{aligned}
  X^{C}=\frac{1}{T \times F}\sum_{i=1}^{T}\sum_{j=1}^{F}X_{ij},
\label{SE_Avg}
\end{aligned}
\end{equation}
where $X^{C}$ is the results of global average pooling of the intermediate feature map $X$ on time and frequency dimensions. Next is to learn the channel-wise attention weights of $X^{C}$. The channel-wise attention mask learning is formulated as: 
\begin{equation}
\begin{aligned}
  M^C = \delta (W_2^{C \times C'} \times (relu((W_1^{C' \times C} \cdot X^C)))),
\label{SE_Formula}
\end{aligned}
\end{equation}
\noindent where $\cdot$ means matrix multiplication. $W_1^{C' \times C}$ and $W_2^{C \times C' }$ are two fully-connected layers which capture the inter-dependence of channels in the intermediate speech feature map $X$. 
The dimensional reduction factor $r=\frac{C}{C'}$ indicates the bottleneck output feature map reduction ratio to avoid the parameters overhead.  Finally, a sigmoid function $\delta$ is used to scale the channel-wise weights. The attention mask is a set of attention factors predicted by supervised speaker classification loss aiming to emphasize essential channels and compress the useless channels. As shown in Fig.~\ref{fig:SE}, the recalibration operation means that the attention mask dot multiplies with the original tensor $X$.

\subsection{Encode Time-Frequency Information}
The above SE attention averages the time-channel-frequency feature map $X^{T \times C \times F}$ into $X^{C}$, which may lose the discriminative speaker information in temporal and frequency domain. Moreover, the SE attention neglects the exploration in global context relationships on time and frequency dimensions. To alleviate these issues, we firstly encode global time and frequency information into the channel-wise feature and then conduct the duality channel-wise attention with the assist of temporal and frequency dimensions respectively. The DTCF attention module is illustrated in Fig.~\ref{fig:details}. We average the time and frequency domains in parallel with 2D adaptive average pooling~(GAP(1D)) as Eq.~\ref{ESE_Avg1} and Eq.~\ref{ESE_Avg2}. GAP(1D) denotes average operation on time or frequency separately. After we concatenate $X^{C \times F}$ and $X^{C \times T}$, a convolution operation with kernel size $(1 \times 1 \times C')$ is adopted to learn the inter-dependence relationships between channels with the help of global time and frequency information. Here $W^{C' \times C}$ is the weights indicating the relationship of global channels with aggregation of temporal and frequency dimensions.

\begin{equation}
\begin{aligned}
    X^{C \times F}=\frac{1}{T}\sum_{i=1}^{T} X_{i},
\label{ESE_Avg1}
\end{aligned}
\end{equation}

\begin{equation}
\begin{aligned}
    X^{C \times T}=\frac{1}{F}\sum_{i=1}^{F} X_{j},
\label{ESE_Avg2}
\end{aligned}
\end{equation}

\begin{equation}
\begin{aligned}
    X_1^{C' \times (T+F)}=relu(W_1^{C' \times C} \cdot [X^{C \times F},X^{C \times T}]),
\label{ESE_Cat}
\end{aligned}
\end{equation}
where $\cdot$ is matrix multiplication. The dimensional reduction factor $r=\frac{C}{C'}$ is used to avoid the exceed of parameters. 

\subsection{Duality Channel-wise Attention}
\begin{figure*}[htp]
  \centering
  \centerline{\includegraphics[width=16cm]{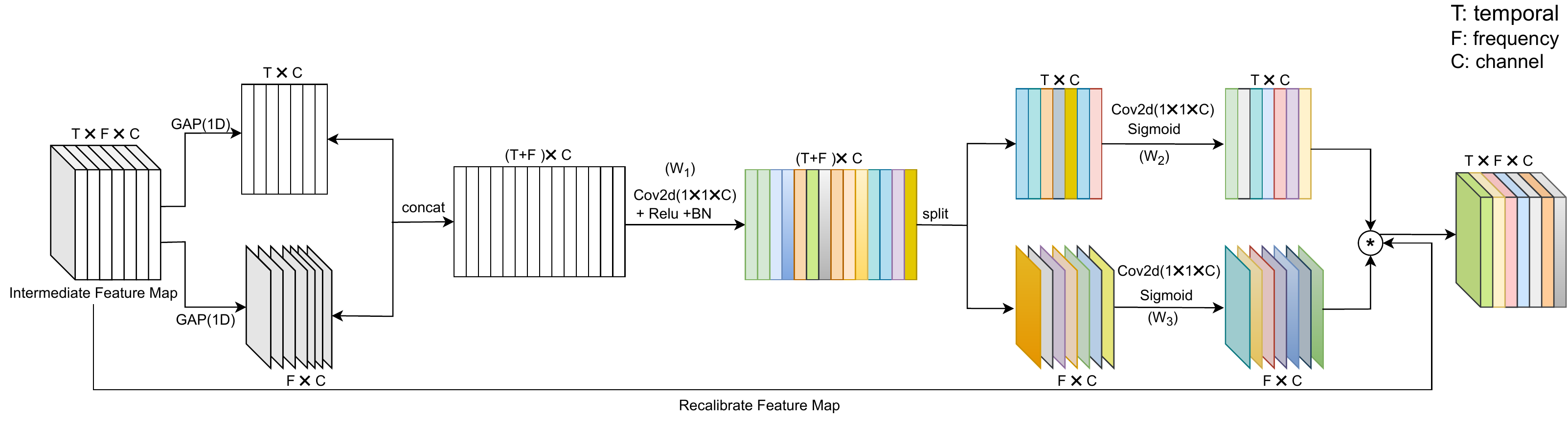}}
 \setlength{\leftskip}{-60pt}
  \setlength{\rightskip}{-60pt}
\caption{Duality Temporal-Channel-Frequency~(DTCF) Attention Block}
\label{fig:details}
\end{figure*}

After encoding the global temporal and frequency domains into the channel attention mask learning, we split $X_1^{C' \times (T+F)}$ into  $X_2^{C' \times T}$ and $X_2^{C' \times F}$ respectively. Then we conduct channel-wise attention twice with the help of temporal and frequency domain separately. The formulas of the duality channel-wise attention are as follows:

\begin{equation}
\begin{aligned}
   X_2^{C \times T} = \delta(W_2^{C \times C'} \cdot X^{C' \times T}),
\label{ESE_FC1}
\end{aligned}
\end{equation}

\begin{equation}
\begin{aligned}
   X_2^{C \times F} = \delta(W_3^{C \times C'} \cdot X^{C' \times F}),
\label{ESE_FC1}
\end{aligned}
\end{equation}
where $\cdot$ is matrix multiplication. $W_2^{C \times C'}$ and $W_3^{C \times C'}$ are channel-wise attention weights with the convolutional operations of the kernel size of $(1 \times 1 \times C')$. $X^{C \times F}$ and $X^{C \times T} $ are the channel-wise attention masks calculated with the help of frequency domain and temporal domain respectively. $\delta$ is sigmoid function to normalize the attention weights into attention masks $W_2^{C \times C'}$ and $W_3^{C \times C'}$. The above proposed attention module is called as Duality Temporal-Channel-Frequency~(DTCF) attention, because it learns duality channel-wise attention with the help of time and frequency information. As shown in Fig.~\ref{fig:details}, the attention masks $X^{C \times F}$ and $X^{C \times T}$ dot multiply the original feature map $X^{T \times C \times F}$ to enhance the necessary channels and suppress the useless channels with consideration of global time and frequency information.

In a word, the proposed DTCF attention learns the global context information of speech feature map with the consideration of both temporal and frequency dimensions. At the same time, the DTCF attention learns the inter-dependence channel relationship just like the SE attention mechanism \cite{zhou2019deep}. 

 
\section{Experimental Setup}
\label{sec:typestyle}
\subsection{Datasets and Data Augmentation}
To evaluate the performance of the DTCF attention, we conduct experiments on two popular datasets CN-Celeb \cite{fan2020cn} and VoxCeleb \cite{nagrani2017voxceleb,chung2018voxceleb2}.

\noindent \textbf{CN-Celeb}
There are 3000 speakers and 11 different genres in CN-Celeb, which involves various real-world noise, cross-channel mismatch, and speaking styles in the wild speech utterances. The training set has more than 600,000 utterances of 2800 speakers and 18,024 utterances of 200 speakers. There are 3,604,800 pairs in the test trials. Moreover, domain mismatch between the enrollment and test in the trials makes this dataset a very challenging one in speaker verification.

\noindent \textbf{VoxCeleb} VoxCeleb2 contains over 1 million utterances from 6,112 speakers extracted from videos uploaded to YouTube. Short-term utterances (7.8s on average) and diverse acoustic environments make it more challenging for SV. We train the speaker verification models with VoxCeleb2 development part, which has in-the-wild speech utterances from 5994 speakers. The evaluation set is Voxceleb1, which contains over 100,000 utterances for 1,251 speakers. There are three types of evaluation trials, which are VoxCeleb1-O, VoxCeleb1-H, and Voxceleb1-E. The VoxCeleb1-O test set is composed of speech from 40 speakers. The extended VoxCeleb1-E list uses the entire VoxCeleb1 set (1251 speakers), and the hard VoxCeleb1-H list contains speakers with the same gender and nationality.

\noindent \textbf{Data Augmentation} The experiments on CN-Celeb and VoxCeleb both do online data augmentation \cite{cai2020fly} during training. The noise datasets used in data augmentation are MUSAN \cite{snyder2015musan} and RIRs \cite{habets2006room}. Moreover, we adopt speed perturbation and specaugment \cite{park2019specaugment} during training. The speed perturbation includes speeding up or down each utterance by 0.9 and 1.1 times respectively. Note that online data augmentation has been successfully applied in various speech and speaker recognition tasks \cite{yamamoto2019speaker,ko2015audio,park2019specaugment}.  

\subsection{Network Configurations}
We validate the proposed DTCF attention module in ResNet34 \cite{heo2020clova}, which only has one-quarter of the channels in each residual block compared with the original ResNet-34~\cite{he2016deep} in order to reduce computational cost. 
We use attention statistic pooling~(ASP)~\cite{heo2020clova} as the pooling layer with the intermediate channel dimension of 128. We adopt the penultimate layer of the network as the embedding extraction layer with 512 dimensions.  The dimensional reduction factor $r$ in attention modules is set to 8. The supervised loss of training we use is Additive Angular Margin Softmax~(AAM-Sotfmax) \cite{deng2019arcface}. The details of the model structure with approximately 9 million parameters are shown in Table~\ref{tab:backbone}. In this paper, SE attention or its variants are embedded into the same ResNet backbone except for the DTCF attention modules (DTCF-Module).

\begin{table}[th]
\centering
\footnotesize
\caption{ResNet34-DTCF Structure }
\label{tab:backbone}
\resizebox{\linewidth}{!}{
\begin{tabular}{lclclcl}
\hline
Layer    & \multicolumn{2}{c}{Kernel Size}               & \multicolumn{2}{c}{Stride}         & \multicolumn{2}{c}{Output Shape}                       \\ \hline
Conv1    & \multicolumn{2}{c}{$ 3\times 3 \times 32 $}   & \multicolumn{2}{c}{$ 1 \times 1 $} & \multicolumn{2}{c}{$ L \times 80 \times 32 $}          \\ \hline
Res1     & \multicolumn{2}{c}{$ 3 \times 3 \times 32$}   & \multicolumn{2}{c}{$ 1 \times 1 $} & \multicolumn{2}{c}{$ L \times 80 \times 32 $}          \\ \hline
DTCF-Module & \multicolumn{2}{c}{-}                         & \multicolumn{2}{c}{-}              & \multicolumn{2}{c}{$ L \times 80 \times 32 $}          \\ \hline
Res2     & \multicolumn{2}{c}{$ 3 \times 3 \times 32 $}  & \multicolumn{2}{c}{$ 1 \times 1 $} & \multicolumn{2}{c}{$ L \times 40 \times 64 $}          \\ \hline
DTCF-Module & \multicolumn{2}{c}{-}                         & \multicolumn{2}{c}{-}              & \multicolumn{2}{c}{$ L \times 40 \times 64 $}          \\ \hline
Res3     & \multicolumn{2}{c}{$ 3 \times 3 \times 64 $}  & \multicolumn{2}{c}{$ 2 \times 2 $} & \multicolumn{2}{c}{$ L_{/2} \times 20 \times 128$} \\ \hline
DTCF-Module & \multicolumn{2}{c}{-}                         & \multicolumn{2}{c}{-}              & \multicolumn{2}{c}{${ L_{/2} \times 20 \times 128 }$} \\ \hline
Res4     & \multicolumn{2}{c}{$ 3 \times 3 \times 128 $} & \multicolumn{2}{c}{$ 2 \times 2 $} & \multicolumn{2}{c}{${L_{/4} \times 10 \times 256 }$}  \\ \hline
DTCF-Module & \multicolumn{2}{c}{-}                         & \multicolumn{2}{c}{-}              & \multicolumn{2}{c}{${L_{/4} \times 10 \times 256 }$}  \\ \hline
Flatten  & \multicolumn{2}{c}{-}                         & \multicolumn{2}{c}{-}              & \multicolumn{2}{c}{${L_{/8} \times 2560 }$}           \\ \hline
ASP      & \multicolumn{2}{c}{-}                         & \multicolumn{2}{c}{-}              & \multicolumn{2}{c}{5120}                               \\ \hline
Linear   & \multicolumn{2}{c}{512}                       & \multicolumn{2}{c}{-}              & \multicolumn{2}{c}{512}                                \\ \hline
AAMSoftmax   & \multicolumn{2}{c}{-}                       & \multicolumn{2}{c}{-}              & \multicolumn{2}{c}{NumSpkrs}                                \\ \hline
\end{tabular}
}
\vspace{-0.5cm}
\end{table}

\subsection{Training Details}
Eighty-dimensional mel-filter bank features with 25ms window size and 10ms window shift are extracted as model inputs. During training, the batch size of every iteration is 300 with 2 GeForce RTX 3090. The learning rate of all model training varies between 1e-8 and 1e-3 using the triangular2 policy \cite{smith2017cyclical}. The optimizer during training is Adam optimizer \cite{kingma2014adam}. The hyperparameter scale and margin of AAM-softmax are set to 30 and 0.2 respectively. To prevent overfitting, we apply a weight decay on all weights in the model of 2e-5. 
\subsection{Scoring Criterion}
In the test phase, we use cosine similarity as the scoring criterion. The performance metrics are equal error rate (EER) and minimum detection cost funtion (minDCF)~\cite{sadjadi20172016} which is evaluated with $P_{target} = 0.01, C_{miss} = C_{fa} = 1$.

\section{Experimental Results and Analysis}
\label{sec:majhead}
\subsection{Results on CN-Celeb Dataset}
The experimental results on CN-Celeb are shown in Table~\ref{tab:eval_cn-celeb}. The EER of the baseline ResNet34 on CN-Celeb is 16.52\%. After embedding the SE attention into each residual block in ResNet34, the EER is reduced to 15.47\% and the minDCF becomes 0.6679. When we replace the SE attention with the proposed DTCF attention module in ResNet34, the EER/minDCF further reduce 0.63\%/0.0718 compared with the results of the ResNet34-SE model.
\begin{table}[hbp]
\caption{Experimental Results on CN-Celeb~(E)}
\resizebox{\linewidth}{!}{
\tiny
\begin{tabular}{lcl}
\hline
 
\multirow{2}{*}{Model} & \multicolumn{2}{c}{CN-Celeb~(E)}   \\ \cline{2-3} 
                       & \multicolumn{1}{l}{EER~(\%)}  & minDCF \\ \hline
ResNet~\cite{li2020cn}              & 16.52                   & -  \\
ResNet34-SE       & 15.47                    & 0.6679  \\
\textbf{ResNet34-DTCF}     & \textbf{14.84} & \textbf{0.5961}
  \\ \hline
\end{tabular}}

\label{tab:eval_cn-celeb}
\end{table}

\begin{table*}[hth]
\centering

\caption{Experimental Results on VoxCeleb}
\resizebox{\linewidth}{!}{
\tiny
\label{tab:voxceleb}
\begin{tabular}{llllllll}
\hline
                                    & \multicolumn{2}{c}{VoxCeleb1-O}                      & \multicolumn{2}{c}{VoxCeleb1-E} & \multicolumn{2}{c}{VoxCeleb1-H}                 \\ \cline{2-7} 
\multirow{-2}{*}{Methods}           & EER~(\%)                                  & minDCF         & EER~(\%)                & minDCF              & \multicolumn{1}{l}{EER~(\%)} & \multicolumn{1}{l}{minDCF} \\ \cline{1-7} 
Ecapa(C=1024)~\cite{desplanques2020ecapa}                       & 0.87                                 & \textbf{0.1066}          & \textbf{1.12}               & 0.1318               & 2.12                    & 0.2101                      \\
ResNet34-ft-CBAM~\cite{yadav2020frequency}                       & 1.08                                    & 0.1655              & 1.43              & 0.1562              &2.67                       & 0.2633                          \\
ResNet34-L2-tf-GTFC~\cite{xia2020speaker}                       & 0.90                                 & 0.1278          & 1.22               & 0.1484               & 2.42                    & 0.2497                      \\
ResNet34-SE                       & 1.15                                & 0.1353          & 1.52               & 0.1641               & 2.83                    & 0.2835                     \\
\textbf{ResNet34-DTCF}                     &  \textbf{0.79}                                & 0.1090          &1.13               & \textbf{0.1259}               & \textbf{2.09}                    & \textbf{0.2082}                    
   \\ \hline
\end{tabular}
}
\vspace{-0.3cm}
\end{table*}
\subsection{Results on VoxCeleb Dataset}
The experimental results on VoxCeleb are shown in Table \ref{tab:voxceleb}. We compare our results with Ecapa \cite{desplanques2020ecapa}, ResNet34-ft-CBAM \cite{yadav2020frequency}, ResNet34-SE~\cite{zhou2019deep}, ResNet34-L2-tf-GTFC \cite{yadav2020frequency}. The results of ResNet34-ft-CBAM, ResNet34-L2-tf-GTFC, ResNet34-SE are reproduced with the same channels of residual blocks in ResNet34. From Table \ref{tab:voxceleb}, the DTCF attention module embedded in ResNet-34 outperforms Ecapa \cite{desplanques2020ecapa} on VoxCeleb1-H. The minDCF of Ecapa on VoxCeleb1-O and EER of Ecapa on VoxCeleb1-E are better than those of our model. Compared with ResNet34-SE which is the most classical speaker verification model structure, ResNet34-DTCF achieves 0.36\%/0.0263 reductions on EER/minDCF on VoxCeleb1-O. The DTCF attention modules make 0.39\%/0.0382 and 0.74\%/0.0753 reductions on EER/minDCF on VoxCeleb1-E and VoxCeleb1-H respectively. Moreover, our attention mechanism is better than other recent competitive attention modules in ResNet34.  

\subsection{Analysis}
\begin{figure}[th]
\footnotesize
\setlength{\abovecaptionskip}{0.1cm}
\setlength{\belowcaptionskip}{-0.3cm}
  \centering
  \includegraphics [width=\linewidth]{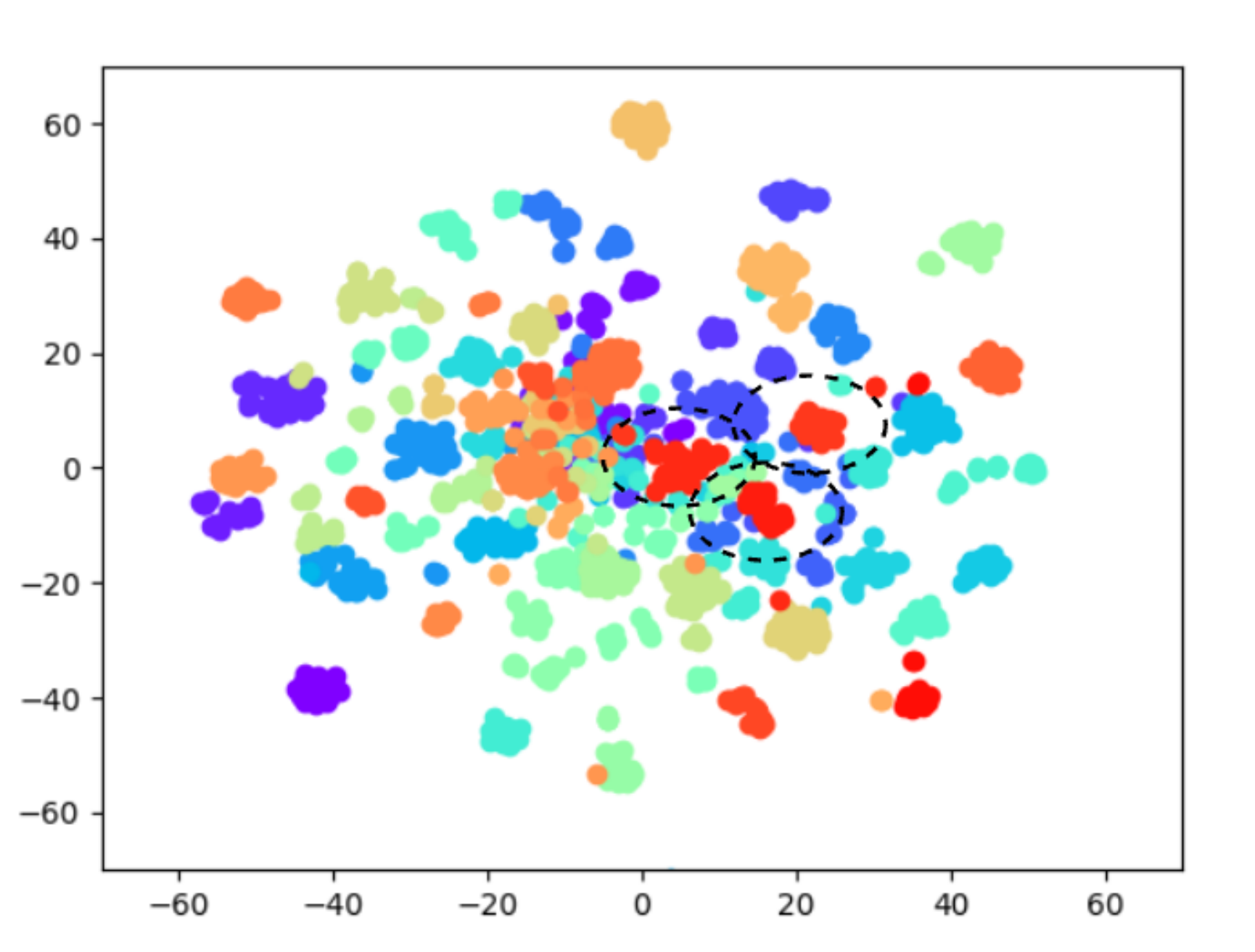}
  \caption{ResNet34-DTCF Embedding Space Visualization}
  \label{fig:analysis0}
\vspace{-0.1cm}
\end{figure}

\begin{figure}[th]
\footnotesize
\setlength{\abovecaptionskip}{0.1cm}
\setlength{\belowcaptionskip}{-0.3cm}
  \centering
  \includegraphics [width=\linewidth]{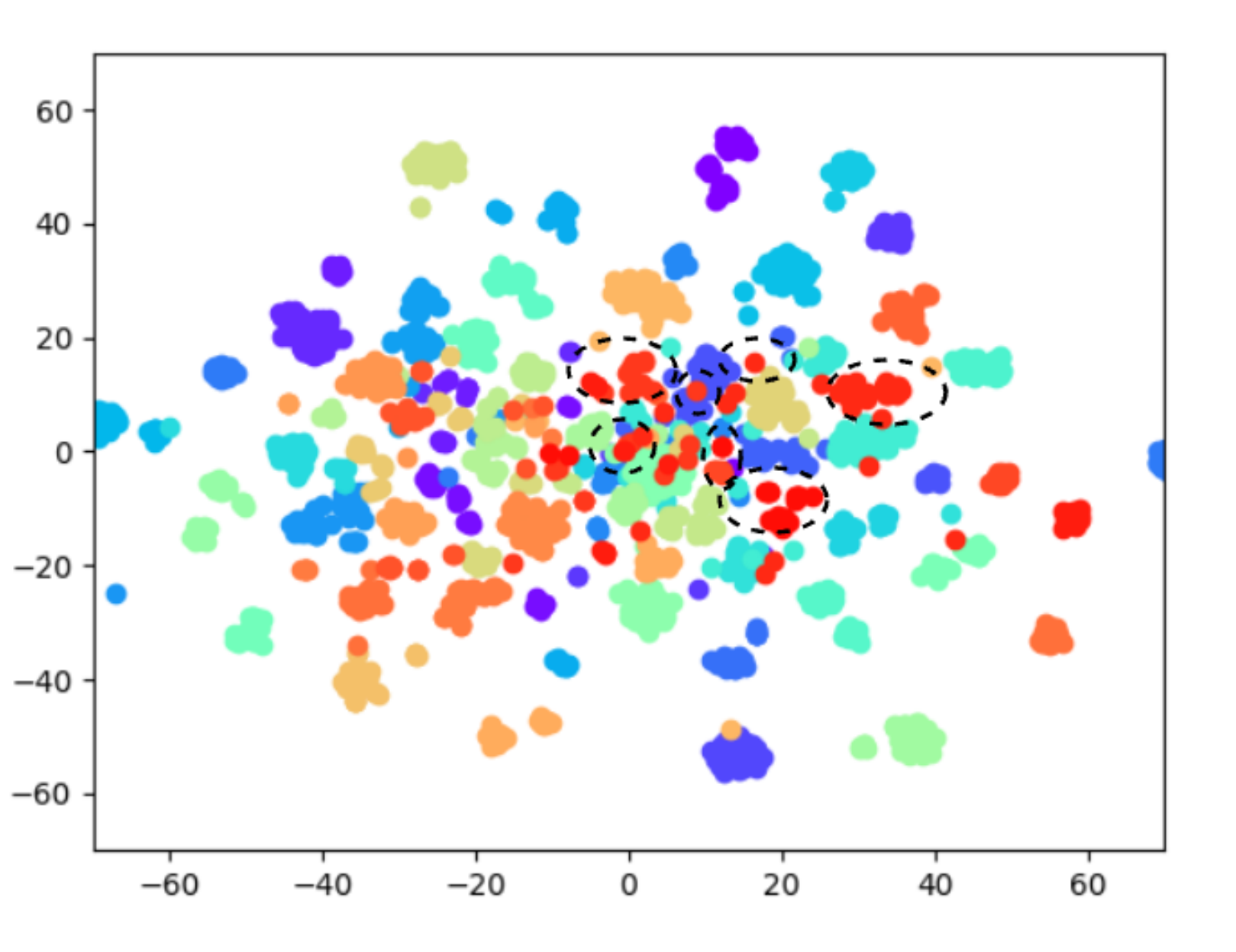}
  \caption{ResNet34-SE Embedding Space Visualization}
  \label{fig:analysis1}
\vspace{-0.1cm}
\end{figure}

We visualize the speaker embeddings of the CN-Celeb evaluation set in Fig.~\ref{fig:analysis0} and Fig.~\ref{fig:analysis1}. Specifically, we randomly select 55 speakers for analysis. One hundred and fifty embeddings from each speaker are randomly chosen for visualization with t-SNE~\cite{van2008visualizing}. Fig.~\ref{fig:analysis0} and Fig.~\ref{fig:analysis1} illustrate the embedding distribution of ResNet34-DTCF and ResNet-SE respectively. Comparing the two figures, we find that the embedding space of ResNet-SE is messier than that of ResNet-DTCF. As shown in the black circle areas in two figures, the speaker category represented by the red dots is clustered into three centers in Fig.~\ref{fig:analysis0}, but there are many scattered points entangled with other categories in Fig.~\ref{fig:analysis1}. This demonstrates that the speaker representation learned from ResNet-DTCF is more discriminative than those learned from ResNet-SE.   

The above experimental results have shown that the proposed DTCF attention module outperforms other channel-wise attention modules both on CN-Celeb and VoxCeleb. We believe that doing simple average pooling along time and frequency domains may lose the necessary information for speaker discriminative representation learning. Instead, our method preserves the time and frequency information when learning the channel-wise attention mask. Specifically, the method can explore the global context in the time and frequency domains as well as make the inter-dependence channel-wise attention mask focus on the salient regions and neglect insignificant parts, leading to more discriminative speaker representation.   

\section{Conclusions}
\label{sec:print}
In this paper, we propose the Duality-Temporal-Channel-Frequency~(DTCF) attention to boost the representation extracting capability of CNN in speaker verification. Different from other channel-wise attention learning with average pooling on time and frequency dimensions, the DTCF attention module firstly recalibrates the channel-wise features with aggregation of global context on temporal and frequency dimensions, and then the duality channel-wise attention is adopted with preservation of temporal and frequency information respectively. 
Experimental results demonstrate that the proposed DTCF attention outperforms other channel-wise attention modules. 
\bibliographystyle{IEEEbib}
\bibliography{Template,refs}

\end{document}